\documentstyle[12pt,epsf,epsfig,psfig]{article}
\newcommand{\gapproxeq}{\lower.7ex\hbox{$\;\stackrel{\textstyle>}{\sim}\;$}}
\newcommand{\lapproxeq}{\lower.7ex\hbox{$\;\stackrel{\textstyle<}{\sim}\;$}}
\begin{document}

\title{Boundaries, Cusps and Caustics\\ in the Multimagnon Continua of \\ 
1D Quantum Spin Systems.}
\author{T.~Barnes \\ \\
{\footnotesize\it Physics Division,
Oak Ridge National Laboratory,} \\
{\footnotesize\it Oak Ridge, TN 37831-6373}  \\
{\footnotesize\it and }  \\
{\footnotesize\it Department of Physics and Astronomy,
University of Tennessee, }\\
{\footnotesize\it Knoxville, TN 37996-1501} \\
}

\vspace{1.5cm}
\maketitle


\begin{abstract}
The multimagnon continua of 1D 
quantum spin systems 
possess several interesting singular 
features that may soon be accessible
experimentally through inelastic neutron scattering. These
include cusps and composition discontinuities 
in the boundary envelopes of two-magnon continuum states and 
discontinuities in the density of states, ``caustics", on and 
within the continuum, which will
appear as 
discontinuities in scattering intensity.
In this note we discuss the general origins of these 
continuum features, and
illustrate our results using the
alternating Heisenberg antiferromagnetic chain
and two-leg ladder as examples. 
\end{abstract}

\maketitle

\eject

\section{Two-magnon states}

\subsection{Introduction and definitions}

Recently the subject of two-magnon excitations 
of quasi-1D quantum spin systems 
has attracted considerable interest. 
Theorists have long predicted that
some of these systems will possess bound states \cite{thyBS}, and there
are now experimental indications of such bound states in 
the alternating chain material 
copper nitrate \cite{Ten00} and in the spin ladder 
(Ca,La)$_{14}$Cu$_{24}$O$_{41}$ \cite{Win01,Gru01}.
\vskip 0.3cm

In addition to the bound modes there is a continuum of two-magnon states,
which has been
reported in inelastic neutron scattering from
polycrystalline Sr$_{0.73}$CuO$_2$
\cite{Mei99} (assuming an alternating chain model)
and
copper nitrate single crystals \cite{Ten00}.
Although this continuum has been considered a rather
uninteresting feature in comparison with 
the bound states, it actually possesses
an interesting and rather complicated internal structure \cite{Tay85,Kar00}. 
With high
resolution inelastic neutron scattering experiments it should be possible
to study these features of the two-magnon continuum in parallel with
studies of the bound modes. To facilitate these experiments, here we
discuss the nature of these interesting features of the continuum and
show how they can be understood simply in terms of aspects of the 
one-magnon modes. 
\vskip 0.3cm

The two-magnon continuum states in a gapped 1D antiferromagnet are
simply composed of two one-magnon excitations, since the interaction
region is a zero matrix element for these unlocalized states. The energy
of a two-magnon state composed of magnons with $k_1$ and $k_2$ is therefore
given by 
\begin{equation}
\omega_2(k = k_1 + k_2) = \omega(k_1) + \omega(k_2)
\end{equation}  
with the usual crystal momentum constraints on $k_1, k_2$ and
$k$, 
\begin{equation}
k = k_1 + k_2  \ mod(2\pi)  \ . 
\end{equation}  
Thus the allowed states in the two-magnon continuum are given by the simple
vector addition of Fig.1.
\vskip 0.3cm

For a fixed two-magnon $k$ we can independently vary $k_1$, keeping
$k_2=k-k_1$. This gives a range of allowed two-magnon energies for given
$k$. Carrying out this ``scan'' in $k_1$ at each $k$ gives the 
full set of states in the two-magnon continuum. 
Interesting physics questions regarding this continuum 
include the determination of the upper- and lower- limit
boundary curves of this allowed region, and the composition of the 
two-magnon states along and within these boundaries.
\vskip 0.3cm

All this information is implicit in the one-magnon dispersion relation
$\omega(k)$. In the following discussion unless otherwise specified 
we will assume that 
$\omega(k)$ is an even, periodic function of $k$ with period $2\pi$, 
monotonically increasing 
inside the range $k=[0,\pi]$, 
with
a single inflection point at $k_0$. We also assume that
$\omega(k)$ departs from its $k=0$ minimum and $k=\pm \pi$ maximum values
quadratically in $k$, and that the inflection point 
lies within $|k|=[0,\pi/2]$).  
(This 
$\omega(k)$ 
is abstracted from the simple alternating chain result,
shown in Fig.1. 
Our discussion can
easily be extended to more general cases.)

\subsection{Lower continuum boundary}

First we consider the lower boundary of the two-magnon continuum, which we call
$\Omega_2^-(k)$.
The $k=0$ two-magnon mode clearly has a lower boundary at 
$\omega_2(0) = 2\omega(0)$, since this satisfies the momentum constraint
$k = k_1 + k_2\ mod(2\pi)$ 
and is the global minimum of free two-magnon energies.
For other $k$ values we wish to solve
\begin{equation}
\Omega_2^-(k) = min_{k_1,k_2=k-k_1} \bigg( \omega(k_1) + \omega(k_2) \bigg)\ .
\end{equation}

Starting from the symmetric state 
$(k_1,k_2) = (k/2,k/2)$, with energy
$\omega_2(k) = 2\omega(k/2)$, we can establish this lower boundary 
$\Omega_2^-(k)$ by
shifting these momenta to $(k_1,k_2) = (k/2-\Delta,k/2+\Delta)$, 
and determining whether any nonzero,
physically distinct
$\Delta$ ($0 < \Delta \leq \pi $) 
gives a lower-energy two-magnon state. (See Fig.2 for this construction.)
The condition that the initial,
symmetric
state itself has the lowest energy is
\begin{equation}
\omega(k/2-\Delta) + \omega(k/2+\Delta) - 2\omega(k/2) > 0 \ \ \forall \ 
\Delta \ ,
\end{equation}
which can be used as the definition of a 
``globally concave" function. 
(A function is locally concave if this is satisfied for
infinitesimal $\Delta$.)
The lower boundary 
curve of the two-magnon continuum is given by
\begin{equation}
\Omega_2^-(k) = 2 \omega(k/2) 
\end{equation}
if the one-magnon dispersion is globally concave.
\vskip 0.3cm

Both upper and lower boundaries will usually be extremal in the energy of
two magnon states, so that they satisfy
\begin{equation}
d\omega_2(k_1,k-k_1) = 0 = \Big( \omega'(k_1) - \omega'(k-k_1) \Big)\, dk_1  
\end{equation}
or equivalently 
\begin{equation}
\omega'(k/2 - \Delta) = \omega'(k/2 + \Delta) \ .
\end{equation}
The search for boundary curves of this extremal type 
can be considered a search for the
minimum- or maximum-energy solutions $\Delta^{\pm}(k)$ of this equation; 
once (7) is solved for
$\Delta^{\pm}(k)$, the boundary curves are
then given by 
\begin{equation}
\Omega_2^{\pm}(k) =  \omega(k/2-\Delta^{\pm}(k)) + \omega(k/2+\Delta^{\pm}(k)) \ .
\end{equation}
Clearly one solution of (7) is $\Delta(k)=0$, giving $\Omega_2(k) = 2\omega(k/2)$.
This may however 
not give the global minimum or maximum two-magnon energy at $k$. 
We shall argue below that the lower boundary is in fact {\it not} given by $\Delta(k)=0$
if $\omega(k/2)$ is convex.
\vskip 0.3cm

If the one-magnon dispersion relation $\omega(k)$ is not
locally 
concave over the full physically independent range $[-\pi,\pi]$, the concave
region $[-k_0,k_0]$ will be bounded by inflection points at $k=\pm k_0$.
For a two-magnon state with momentum $|k|> 2k_0$  
(which lies in the convex region of the one-magnon dispersion relation) 
the lower boundary of the two-magnon continuum is 
not given by $2\omega(k/2)$, since an infinitesimal
perturbation into an
unsymmetric state $(k_1,k_2) = (k/2-\Delta,k/2+\Delta)$
lowers the energy;
\begin{equation}
\lim_{\Delta \to 0}  
\omega(k/2-\Delta) + \omega(k/2+\Delta) - 2 \omega(k/2) 
= \Delta^2 \omega''(k/2) + O(\Delta^4) < 0 \ .
\end{equation}

The 
nature of the departure of $\Omega_2^-(k)$ from the symmetric state 
as we enter the convex region depends on the behavior of 
$\omega(k)$ near the inflection point $k_0$. 
If the departure from the symmetric state $\Delta(k)=0$ is continuous,
$\Delta(k)$ will be small near $k=2k_0$, and we can expand (7) to find
\begin{equation}
\lim_{k\to {2k_0}_+} \Delta(k)  = 
\bigg[ {-6 w^{(ii)}(k/2)  \over w^{(iv)}(k/2)} \bigg]^{1/2} \ .
\end{equation}
We can further simplify this by expanding in $k-2k_0$ and using $\omega''(k_0)=0$, which gives
\begin{equation}
\lim_{k\to {2k_0}_+} \Delta(k)  = 
\bigg[ {-3 w^{(iii)}(k_0) \over w^{(iv)}(k_0)} \bigg]^{1/2}
\Big( k - 2k_0 \Big)^{1/2} \ .
\end{equation}
Thus if the fourth derivative 
$\omega^{(iv)}(k_0)$ is positive 
($\omega^{(iii)}(k_0) < 0$ at this inflection point),
the minimum energy state 
$(k_1,k_2) = (k/2-\Delta(k),k/2+\Delta(k))$
departs from the symmetric one $(k/2,k/2)$ as a square root of momentum.
\vskip 0.3cm

The corresponding departure in energy 
of $\Omega_2^-(k)$
from $2\omega(k/2)$ 
is much less abrupt, and for small $k-2k_0$ is
\begin{equation}
\lim_{k\to {2k_0}_+} \Omega_2^-(k) - 2\, \omega(k/2)\  = 
\ -{3\over 4}\; \bigg[ {w^{(iii)}(k_0)^2 \over w^{(iv)}(k_0)} \bigg]
\Big( k - 2k_0 \Big)^2 \ .
\end{equation}

This piecewise continuous behavior and square-root 
departure in composition
$(k_1,k_2)$ from symmetric states on $\Omega_2^-(k)$
near $2k_0$ is illustrated by the alternating chain in Fig.3.
In this model to leading order in
$\alpha$ (see Appendix) one has
$k_0 - \pi/2 = - \alpha/2$, $w^{(iii)}(k_0)= - \alpha/2$ and
$w^{(iv)}(k_0) = 3 \alpha^2/4$, so
\begin{equation}
\lim_{k\to {2k_0}_+\atop \alpha\to 0}
\Delta(k)  =
\bigg( {2\over \alpha }\bigg)^{1/2}
\Big( k - (\pi - \alpha) \Big)^{1/2} 
\end{equation}
and
\begin{equation}
\lim_{k\to {2k_0}_+\atop \alpha\to 0}
\Omega_2^-(k) - 2\, \omega(k/2)  =
\ -{1\over 4}\; 
\Big( k -  (\pi - \alpha) \Big)^2 \ .
\end{equation}
The location of this composition discontinuity could be used
as an independent experimental determination of $\alpha$ that does not
require following $\omega(k)$ over a wide range of $k$. 
\vskip 0.3cm

Between $|k| = 2k_0$ and $|k|=\pi$ the two-magnon lower boundary 
passes through states of increasingly unsymmetrical composition, reaching 
the limit $k_1=0, |k_2|=\pi$ at $|k|=\pi$. This is illustrated in Fig.3.
\vskip 0.3cm

Were $w^{(iv)}(k_0)$ 
negative we would find a discontinuous jump in the value of $k_1$ 
on the boundary $\Omega_2^-(k)$ at $k=2k_0$. In either case we expect to 
see rapid variation in scattering intensities from states near
this composition discontinuity.

\subsection{Upper continuum boundary}

The upper two-magnon continuum boundary, defined by
\begin{equation}
\Omega_2^+(k) = max_{k_1,k_2=k-k_1} \bigg( \omega(k_1) + \omega(k_2) \bigg)\ ,
\end{equation}
can be found using a similar 
construction to Fig.2.
First note that the global maximum
of energy of two-magnon states is $2\omega(\pi)$, 
at $k=2\pi$ (equivalent to $k=0$ or any $k=2\pi m$)
and consists of a symmetric state of two 
magnons,
$k_1=k_2=\pi$.
As we move away from this highest-energy state at $k=2\pi$, 
we can form analogous 
symmetric states with
total momentum $2\pi - k$ (equivalent to $-k$ and degenerate
with a $k$ state) from two single-magnon states with
$k_1 = k_2 = \pi - k/2$, as shown in Fig.4. Evidently this state
has energy $2 \omega(\pi - k/2)$.
We can determine whether this
is the maximum energy state at this $k$ 
by considering a symmetrically displaced 
combination $(k_1, k_2) = (\pi - k/2 - \Delta,  \pi - k/2 +\Delta)$,
analogous to Fig.2. The 
energy difference is
\begin{equation}
\Delta E = 
\omega(\pi - k/2+\Delta) + \omega(\pi - k/2-\Delta) - 2 \omega(\pi - k/2) \ ,
\end{equation}
which is negative for all $k$ in $[0,\pi]$ 
(meaning that the symmetric state has the highest energy) 
if $\omega(k)$ is convex over the range $[\pi/2, \pi ]$. 
If this is satisfied the upper boundary curve of the two-magnon continuum 
is given by
\begin{equation}
\Omega_2^+(k) = 2 \omega(\pi - k/2) \ .
\end{equation}
This is the case for our alternating chain example, as is shown in Fig.4.
It is also satisfied for the uniform $S=1/2$ Heisenberg chain, for which
$\omega(k) = (\pi / 2) J\, |\sin(k)|$; since this is a convex
function, (17) implies
\begin{equation}
\Omega_2^+(k) = 2 \omega(\pi - k/2) = \pi J\,|\sin(k/2)| \ ,
\end{equation}
which is the well-known upper boundary of the two-spinon continuum
\cite{Des62}.
\vskip 0.3cm

If instead we encounter an inflection point $\omega''(k_0)=0$ in 
the interval $[\pi/2, \pi ]$,
we will find a departure from 
$2 \omega(\pi - k/2) $
along the upper boundary. 
As in Fig.3 this will be accompanied by a change
in composition from the symmetric state $(k_1,k_2)=(\pi - k/2,\pi - k/2)$.
\vskip 0.3cm

Since we have 
assumed for our example that the inflection point $k_0$ lies in
the interval $[0, \pi/2 ]$, we do not encounter an inflection 
point, and the 
function $\Omega_2^+(k) = 2 \omega(\pi - k/2)$ describes the entire
upper boundary curve for $k\in [-\pi,\pi]$. This function is monotonically
decreasing inside the range $k=[0,\pi ]$, and 
has a negative slope at $k=\pi$, given our assumption
that $\omega(k)$ has a positive slope at $\pi / 2$.
Reflection 
symmetry of the dispersion relation about
$k=\pi$ then implies a cusp in the slope of the upper boundary
curve at $k=\pi$, as shown in Fig.4.
\vskip 0.3cm

In Fig.5 we show the complete result for the boundary curves
$\Omega_2^+(k)$ and  $\Omega_2^-(k)$ of the two-magnon
continuum for our alternating chain example.
Note that the two-magnon bandwidth minimum at $k=\pi$ is not zero; 
the value 
\begin{equation}
W_2^{min} = 
\Omega_2^+(\pi) - \Omega_2^-(\pi) =
2 \omega(\pi/2) - \omega(0) - \omega(\pi) \ ,
\end{equation}  
is $\alpha^2/4 + \alpha^3/8  + O(\alpha^4)$ 
for the alternating chain. A zero bandwidth at $k=\pi$ would follow 
for example from pure 
$\cos(k)$ modulation in the one-magnon dispersion relation.
\vskip 0.3cm

\subsection{More general boundary curves}

Motivated by the alternating chain as a prototypical gapped 1D 
quantum spin system, we have
assumed that the one-magnon dispersion relation is even, monotonically 
increasing inside  $[0, \pi ]$, and has a single
inflection point at $k_0$ inside $[0, \pi/2 ]$. 
This led to an upper two-magnon 
continuum boundary with a cusp and a symmetric composition $k_1=k_2$ everywhere,
and a lower two-magnon 
boundary with a composition break at $2k_0$. 
\vskip 0.3cm

Although these results are
valid in general for alternating chains, in other quantum spin systems 
we may encounter a single inflection point in $[\pi /2, \pi ]$, 
or a different
number of inflection points. 
An example of more complicated 
behavior is provided by the two-leg ladder, which 
for sufficiently large $\alpha$
has a one-magnon
dispersion relation $\omega(k)$ with a 
minimum at $k=\pi$, a
secondary minimum at
$k=0$, 
two inflection points in $k \in \; [0,\pi ]$,
and a maximum at an intermediate $k \in \; [0, \pi/2 ]$. 
\vskip 0.3cm

We expect to find 
a symmetric state $k_1=k_2=k/2$ at the lower boundary 
$\Omega_2^-(k)$ of the continuum
when $\omega(k/2)$ is
globally concave, and an
unsymmetric two-magnon state on $\Omega_2^-(k)$ with an energy
$ < 2 \omega(k/2)$
when $\omega(k/2)$ is a locally or globally convex function of $k$. 
With more than one inflection point, 
boundary curves will typically be composed of pieces with symmetric and 
asymmetric composition.

\subsection{Density of states: caustics}

Our results for the composition of two-magnon states along the upper and
lower continuum boundaries imply the existence of discontinuities 
in the density of states on the continuum boundary and under certain
conditions within the continuum. We refer to any such discontinuity
in the density of states as a ``caustic".
\vskip 0.3cm

Consider the density of states in a system of $N$ spins
with unit cell $b$; the number of one-magnon states $dn$ in an interval 
$dk$ is
\begin{equation}
dn = \bigg( {2\pi \over N b} \bigg)\; dk \equiv n_0 \; dk \ .
\end{equation}
and for two-magnon states in $dk_1dk_2$ it is
\begin{equation}
d^{\, 2}n  = n_0^2 \; dk_1 dk_2 \ .
\end{equation}

Since we observe two-magnon continuum states with specified 
total energy $E=\omega(k_1)+ \omega(k_2) + E_0$ and total momentum 
$k=k_1+k_2$, the number of states $d^{\, 2}n$ 
in a region $dk\, dE$ is more relevant
to experiment. This density has a Jacobean factor,
\begin{equation}
d^{\, 2}n  =  
{n_0^2 \over |\, \omega'(k_1) - \omega'(k_2)\, | } 
\, dk\, dE \equiv \rho(k,E)\, dk\, dE \ .
\end{equation}
Note that the condition for a divergent density of states $\rho(k,E)$, 
which is a singular caustic,
is
\begin{equation}
\omega'(k_1) =  \omega'(k_2) \ ,
\end{equation}
which was exactly the condition (7) 
used to search for boundaries of the two-magnon
continuum. Thus we expect to find a divergent density of states $\rho(k,E)$ on
the two-magnon boundary curves $\Omega_2^-(k)$ and $\Omega_2^+(k)$.
\vskip 0.3cm

Unfortunately we cannot invert the 
$(E,k)\leftrightarrow (k_1,k_2)$ relationship
and solve (22) for $\rho(k,E)$ for a general one-magnon
dispersion relation $\omega(k)$. We can however
determine this density of two-magnon states in certain limits, including the
case of states close to the boundaries.
\vskip 0.3cm

First we consider the symmetric lower boundary 
$\Omega_2^-(k) = 2\omega(k/2)$. 
(Recall that $2\omega(k/2)$ gives the lower boundary of the
two-magnon continuum for a range of $k$, as discussed 
in previous sections.)
Perturbing the momenta away from this 
symmetric point into the continuum of
asymmetric states, we can invert $(E,k)(k_1,k_2)$ infinitesimally
and use (22) 
to find the density of states $\rho(k,E)$ near
$2\omega(k/2)$. This gives 
\begin{equation}
\lim_{E\to 2\omega(k/2)} \rho(k,E) = {n_0^2 \over 2\, |\, \omega''(k/2)\, |^{1/2} } \
\bigg|  E - 2\omega(k/2) \bigg|^{-1/2} \ .
\end{equation}
which evidently has a $1/\sqrt{\Delta E}$ singularity as we approach
$2\omega(k/2)$.
This behavior is reminiscent of the van Hove singularity in the 
density of states as we approach a band edge, 
although that single particle 
density is instead proportional to 
$1/|\omega'(k)|$.  
\vskip 0.3cm

The asymmetric region of $\Omega_2^-(k)$ with $k_1 < k_2$ 
also has a singular density of states,
since $\omega'(k_1) = \omega'(k_2)$ there as well. We may again solve for 
$\rho(k,E)$ 
near the boundary 
using (22), with the result
\begin{equation}
\lim_{E\to \omega(k_1) + \omega(k_2)} \rho(k,E) = 
{n_0^2 \over |\, 2\, (\omega''(k_1) + \omega''(k_2))\, |^{1/2} } \
\bigg|  E - \omega(k_1) -  \omega(k_2)\bigg|^{-1/2} 
\end{equation}
where $(k_1,k_2) = k/2 \mp \Delta(k)$. The symmetric boundary 
formula (24) is a special case of this more general result.
\vskip 0.3cm

A final special case is the density of states in $E$ along the line $k=2k_0$; since
$\omega''(k_0)=0$ this is a singular case. Expanding (7) to $O(\Delta^3)$ (the 
usual leading $O(\Delta)$ term vanishes), we find
\begin{equation}
\lim_{E\to 2\omega(k_0)} \rho(k=2k_0,E) 
= 
n_0^2\; \bigg[ {3\over 4\, |\, \omega^{(iv)}(k_0)\, |}\bigg]^{1/4}  \
\bigg|  E - 2\omega(k_0) \bigg|^{-3/4} \ .
\end{equation}

Although we have specifically discussed the singular density of states 
found at the upper and
lower boundaries $\Omega_2^{\pm}(k)$, we note that there may
be caustics
specified by (7) {\it within} the continuum as well, 
and that these will also have
square-root divergences in $\rho(k,E)$ as we approach the singular line. 
\vskip 0.3cm

We again use the alternating chain to illustrate this. 
In Fig.6 we show points in the 
two-magnon continuum, generated with a $\rho(k,E)$ distribution. 
The divergences
in $\rho(k,E)$ on the upper and lower boundaries, 
``border caustics", are evident (see also Fig.10).
Fig.7 shows an enlargement including 
the asymmetric region $k > 2k_0$; 
a caustic {\it within} the continuum due to the 
solution
$\omega_2(k) = 2\omega(k/2)$ of (7) is apparent.   

\subsection{Spin ladder}

As an independent example of a two-magnon continuum, 
in Fig.8 we show the one-magnon dispersion $\omega(k)$
and two-magnon continuum for a two-leg spin ladder, as in Fig.6 for the
alternating chain. The coupling strengths are 
$\alpha \equiv J_{||} / J_{\perp} = 0.3$ and $J_{\perp}=1$, and the
analytic results used are summarized in the appendix. 
\vskip 0.3cm

Although the ladder one-magnon 
$\omega(k)$
has the locations of maxima and minima translated by $\pi$ 
relative to the alternating chain, the two-magnon continua
of the two systems are qualitatively remarkably similar
for small $\alpha$. This is
because a shift in $k$ by $2\pi$, for two magnons, is equivalent to zero. 
Close inspection reveals that the $\alpha$-dependent effects are
typically rather larger on the ladder than the alternating chain,
presumably because the ladder dimers are coupled by twice as 
many links as the alternating-chain dimers. 

\section{Higher multimagnon states}

\subsection{Definitions}

Higher multimagnon continua show features similar to the cusps and
discontinuities of two-magnon states, although these will presumably
be of much less experimental interest due to the weaker coupling of
higher-lying states to experimental probes. We shall nonetheless sketch
the generalization of some of the simpler results for two-magnon 
systems to the $n$-magnon continua.
\vskip 0.3cm

The energy of an $n$-magnon state is given by
\begin{equation} 
\omega_n(k) = \sum_{m=1}^n \omega(k_m)
\end{equation}
where 
\begin{equation} 
k = \sum_{m=1}^n k_m \ mod\ (2\pi) \ .
\end{equation}

Again assuming a gapped one-magnon dispersion relation with the general
features discussed in the previous section (an even $\omega(k)$,
monotonically increasing 
inside $[0,\pi]$, quadratic behavior in $k$ near these endpoints,
and a single inflection point $k_0$), we can 
infer several features of the $n$-magnon continuum.

\subsection{N-magnon continuum boundaries}

Provided that the individual 
magnon momenta $k/n$ are below the inflection point $k_0$, 
($k < nk_0 \ \forall \ k \in [0,\pi]$), 
the lower boundary 
$\Omega_n^-(k)$ 
of the n-magnon continuum 
is given by the energy of 
the symmetric state
$(k_1,k_2\dots k_n) = (k/n,k/n\dots k/n)$, 
\begin{equation}
\Omega_n^-(k) = n \omega(k/n) 
\end{equation}
This implies that for sufficiently large $n$ 
($n > \pi / k_0 $) all the lower-boundary curves $\{ \Omega_n^-(k)\}$
possess cusps at $k=\pi$. Since the departure of $\omega(k)$
from $\omega(0)$ for small $k$ is assumed to be of the form
$\omega(k) = \omega(0) + c_2 k^2 + O(k^4)$, 
the lower boundary curve 
will approach 
\begin{equation} 
\lim_{n\to \infty} \Omega_n^-(k) = 
n\omega(0) + (c_2/n) k^2 + O(n^{-3}) \ . 
\end{equation}
Thus $\Omega_n^-(k)$
asymptotically approaches the line $n\omega(0)$, with a 
residual $O(n^{-1})$ parabolic 
component
$\propto k^2/n$. 
\vskip 0.3cm

The upper boundary $\Omega_n^+(k)$ differs for even and odd magnon number. 
For an even number of magnons the upper boundary maximum energy
$n\omega(\pi)$ occurs at $k=n\pi \equiv 0$. Again assuming quadratic
behavior near the extremum,
$\omega(k) = \omega(\pi) - \tilde c_2 (\pi - k)^2 
+ O((\pi - k)^4)$, 
the symmetric solution 
$(k_1,k_2\dots k_n) = (\pi - k/n, \pi - k/n\dots \pi - k/n)$ with
energy
\begin{equation}
\Omega_n^+(k) = n\omega(\pi - k/n) 
\end{equation}
gives the upper boundary curve provided that we do not encounter
an inflection point, $\pi - k/n > k_0$. Once again this will be satisfied 
for sufficiently large $n$, and all these even-$n$ upper boundary curves 
have a cusp at $k=\pi$. The behavior of this upper boundary for 
large $n$ approaches a constant with a negative parabolic 
component of $O(n^{-1})$, 
\begin{equation} 
\Omega_n^+(k) = 
n\omega(\pi) - (\tilde c_2/n) k^2 + O(n^{-3}) \ , 
\end{equation}
analogous to the result for the lower boundary.
The upper boundary for odd magnon number $n$ differs in that the 
maximum energy state has total 
momentum $k=\pi$ rather than $k=0$, and a cusp at $k=0$ rather than
$k=\pi$.   

\subsection{Caustics}

The higher multimagnon continua also 
possess discontinuities in their densities
of states $\{ \rho(k,E)\} $, 
although these are qualitatively different from the
results we have shown for the two-magnon case. This is largely due 
to the fact that the $(k_1,k_2)\to (k,E)$ mapping gives a simple Jacobean
(22) which leads to caustics where it is singular, whereas the $n\geq 3$
multimagnon continua project a flat distribution in $(k_1,k_2,\dots k_n)$ 
onto a smaller number of variables $(k,E)$. This involves an integration
over $n-2$ variables, which smoothes the singularities seen in the two-magnon
case.
\vskip 0.3cm

As examples of these higher continua, 
in Fig.9 we show the three-magnon continuum for 
our alternating chain example, and Fig.10 shows the density of
states encountered in a fixed-$k$ slice (at $k=\pi/2$) through the
two-, three-, and four-magnon continua. 
(This figure is a histogram of points falling within
$ 0.495\pi < k < 0.505\pi $,  
with an energy binning of 
$\delta E = 0.01$, normalized 
as $f(k,E) = N(k,E)/\delta E \, \delta k \, N_{tot.}$. 
The 
binned points 
were selected from an initial flat distribution, $-\pi < k_1,k_2\dots k_n 
< \pi$,
of $N_{tot.} = 2^{28}$ points.)
\vskip 0.3cm

The two-magnon continuum
in Figs.9,10 clearly shows the $1/\sqrt{\Delta E}$ 
singular border caustics in the density of states,
as given by (24).
In contrast, the three-magnon continuum 
density $\rho(k,E)$ has border caustics with only finite
discontinuities in the density of states,
and the four-magnon
$\rho(k,E)$ is continuous on the boundaries.

\section{Summary and Conclusions}

We have considered the multimagnon continua of 1D
quantum spin systems, and have illustrated our results
using the Heisenberg alternating chain and 
two-leg spin ladder as examples. 
\vskip 0.3cm

The boundaries of the two-magnon continua that result from 
one-magnon dispersion relations similar to the alternating chain case
are considered in detail, and general results are derived for these
and higher multimagnon
boundary curves. It is shown that these curves exhibit cusps
and discontinuous changes in composition under certain conditions.
The density of states 
$\rho(k,E)$ in the
continuum is also considered, and it is noted that $\rho(k,E)$ can 
possess discontinuities and divergences
on ``caustic" lines, both on the continuum boundaries 
and within the continuum. 
\vskip 0.3cm

We anticipate that it should be possible to identify these
features of two-magnon continua through 
high-resolution inelastic neutron scattering
experiments on candidate alternating chain and ladder
materials, and that precise studies of these features may allow
determinations of the parameters 
of spin Hamiltonians through the 
study of relatively small ranges of $(k,E)$
space. 

\section{Acknowledgements}

It is a pleasure to thank B. Lake, G.I. Meijer, G. M\"uller, S.E. Nagler, 
T. Papenbrock, J. Riera,
R.R.P. Singh, D.A. Tennant and G.S. Uhrig for useful discussions 
and communications.
This research was sponsored by 
the Laboratory Directed Research and Development Program of
Oak Ridge National Laboratory (ORNL), 
managed by UT-Battelle, LLC for the U.S.
Department of Energy under Contract No. DE-AC05-00OR22725,
and by
the Department of Physics and Astronomy,
the Neutron Sciences Consortium, 
and
the Chemical Physics Program 
of the University of Tennessee.

\newpage

\renewcommand{\theequation}{A\arabic{equation}}
\setcounter{equation}{0}  
\section*{Appendix: results for alternating chains and ladders}  
\vskip 0.5cm
\hskip 0.8cm
In this appendix we collect some results for the 
alternating Heisenberg chain and two-leg ladder, which we used to illustrate 
aspects of multimagnon continuum states in the paper.
\vskip 0.3cm

\vskip 0.5cm
The alternating chain Hamiltonian is defined by
\begin{equation}
H =   
J \sum_{i=1}^{N/2}
\; {\vec S_{2i-1}} \cdot {\vec S_{2i}}
+ \alpha  \;  {\vec S_{2i}} \cdot {\vec S_{2i+1}} \ ,
\label{Ham}
\end{equation}
where we assume cyclic boundary conditions.
The corresponding one-magnon dispersion relation $\omega(k)$ can be written as a
Fourier series, with $\alpha$-dependent coefficients;
\begin{equation}
\omega(k)
=
\sum_{\ell=0}^\infty a_{\ell}(\alpha)\, \cos(\ell k) \ ,
\end{equation}
where we have implicitly set the strength $J$ and unit cell $b$ to unity.
The Fourier coefficients may be evaluated as a power series in
$\alpha$ using a strong-coupling expansion. The $O(\alpha^5)$ series
of Ref.\cite{Bar99}, which we use to generate 
the numerical results given in this paper, are
\vskip 0.3cm

\begin{equation}
\begin{array}{llrrrrrr}
a_0
& =
& 1
&
& - {1\over 16}\, \alpha^2
& + {3\over 64}\, \alpha^3
& + {23\over 1024}\, \alpha^4
& - {3\over 256}\, \alpha^5
\\
\\
a_1
& =
&
& - {1\over 2}\, \alpha
& - {1\over 4}\,\alpha^2
& + {1\over 32}\,\alpha^3
& + {5\over 256}\, \alpha^4
& - {35\over 2048}\, \alpha^5
\\
\\
a_2
& =
&
&
& - {1\over 16}\,\alpha^2
& - {1\over 32}\,\alpha^3
& - {15\over 512 }\, \alpha^4
& - {283\over 18432}\, \alpha^5
\\
\\
a_3
& =
&
&
&
& - {1\over 64}\,\alpha^3
& - {1\over 48}\, \alpha^4
& - {9\over 1024}\, \alpha^5
\\
\\
a_4 
& =
&  
&  
&  
&  
& - {5\over 1024}\, \alpha^4
& - {67\over 9216}\, \alpha^5
\\
\\
a_5 
& =
&  
&  
&  
&  
& 
& - {7\over 4096}\, \alpha^5
\end{array}
\ .
\label{magnondisp_ach}
\end{equation}
\vskip 0.3cm
We may use these coefficients to develop series for several of the quantities
discussed here. In particular, the inflection point $k_0$ of $\omega(k)$
is given by
\begin{equation}
k_0 = 
{\pi \over 2} 
-{1\over 2}\,\alpha
-{29\over 96}\,\alpha^3
-{119\over 576}\,\alpha^4
+O(\alpha^5) 
\end{equation}
and the energy $\Omega_2^-(2k_0)$ at 
the ``break point" $2k_0$ (see Fig.3) on the lower boundary of the 
two-magnon continuum is given by
\begin{equation}
\Omega_2^-(2k_0) = 2\omega(k_0) =
2
-{1\over 2}\,\alpha^2
-{3\over 32}\,\alpha^3
-{11\over 64}\,\alpha^4
-{311\over 1024}\,\alpha^5
+O(\alpha^6) \ .
\end{equation}
The derivatives of the 
one-magnon energy $\omega(k)$ at $k_0$ determine
the nature of the asymmetric two-magnon lower boundary state for
$k>2k_0$. To $O(\alpha^5)$ these are
\begin{equation}
\begin{array}{lllllll}
\omega'(k_0)
& =
& + {1\over 2}\, \alpha
& + {1\over 4}\, \alpha^2
& - {1\over 64}\, \alpha^3
& - {13\over 256}\, \alpha^4
& + {229\over 4096}\, \alpha^5
\\
\\
\omega''(k_0)
& =
& 0
& 
& 
& 
& 
\\
\\
\omega^{(iii)}(k_0)
& =
& - {1\over 2}\, \alpha
& - {1\over 4}\, \alpha^2
& + {1\over 64}\, \alpha^3
& + {93\over 256}\, \alpha^4
& - {925\over 4096}\, \alpha^5
\\
\\
\omega^{(iv)}(k_0)
& =
& 
& + {3\over 4}\, \alpha^2
& + {3\over 8}\, \alpha^3
& + {63\over 128}\, \alpha^4
& + {257\over 512}\, \alpha^5
\end{array}
\ .
\label{wderivs_ach}
\end{equation}
\vskip 0.3cm

We also give the corresponding results for the
two-leg spin ladder (to one higher order in $\alpha$). 
The $N$-rung ladder Hamiltonian is given by
\begin{equation}
H =   
J_{\perp} \sum_{i=1}^{N}
\; {\vec S_{i,1}} \cdot {\vec S_{i,2}}
+ \alpha  \;  
(
{\vec S_{i,1}} \cdot {\vec S_{i+1,1}} 
+
{\vec S_{i,2}} \cdot {\vec S_{i+1,2}}) \ ,
\label{Ham_lad}
\end{equation}
and the
$\omega(k)$ one-magnon Fourier coefficients 
to $O(\alpha^6)$ are:
\vskip 0.3cm

\begin{equation}
\begin{array}{llrrrrrrr}
a_0
& =
& 1
&
& + {3\over 4}\, \alpha^2
& + {3\over 8}\, \alpha^3
& - {13\over 64}\, \alpha^4
& - {5\over 8}\, \alpha^5
& - {1\over 2}\, \alpha^6
\\
\\
a_1
& =
&
& + \alpha
& 
& - {1\over 4}\,\alpha^3
& - {5\over 16}\, \alpha^4
& - {13\over 64}\, \alpha^5
& + {3\over 32}\, \alpha^6
\\
\\
a_2
& =
&
&
& - {1\over 4}\,\alpha^2
& - {1\over 4}\,\alpha^3
& - {1\over 32 }\, \alpha^4
& + {13\over 64}\, \alpha^5
& + {11\over 64}\, \alpha^6
\\
\\
a_3
& =
&
&
&
& + {1\over 8}\,\alpha^3
& + {1\over 8}\, \alpha^4
& - {3\over 32}\, \alpha^5
& - {81\over 256}\, \alpha^6
\\
\\
a_4 
& =
&  
&  
&  
&  
& - {5\over 64}\, \alpha^4
& - {3\over 32}\, \alpha^5
& + {73\over 1024}\, \alpha^6
\\
\\
a_5 
& =
&  
&  
&  
&  
& 
& + {7\over 128}\, \alpha^5
& + {5\over 64}\, \alpha^6
\\
\\
a_6 
& =
&  
&  
&  
&  
& 
& 
& - {21\over 512}\, \alpha^6
\end{array}
\ .
\label{magnondisp_lad}
\end{equation}
\vskip 1cm
\noindent
The inflection point $k_0$ is:
\begin{equation}
k_0 = 
{\pi \over 2} 
+\alpha
+\alpha^2
+{2\over 3}\,\alpha^3
-{1\over 2}\,\alpha^4
-{1171\over 320}\,\alpha^5
+O(\alpha^6) 
\end{equation}
\vskip 0.3cm
\noindent
and the lower boundary energy $\Omega_2^-(2k_0)$ at 
the break point $2k_0$ is:
\begin{equation}
\Omega_2^-(2k_0) = 2\omega(k_0) =
2
-{3\over 4}\,\alpha^3
-{5\over 4}\,\alpha^4
-{7\over 32}\,\alpha^5
+{3307\over 512}\,\alpha^6
+O(\alpha^7) \ .
\end{equation}

\eject
\noindent
Finally, the derivatives of 
$\omega(k)$ at $k_0$ are:
\vskip 0.3cm
 
\begin{equation}
\begin{array}{llllllll}
\omega'(k_0)
& =
& -  \alpha
& 
& + {1\over 8}\, \alpha^3
& - {5\over 16}\, \alpha^4
& - {117\over 128}\, \alpha^5
& - {223\over 256}\, \alpha^6
\\
\\
\omega''(k_0)
& =
& 0
& 
& 
& 
& 
\\
\\
\omega^{(iii)}(k_0)
& =
& +  \alpha
& 
& - {1\over 8}\, \alpha^3
& + {53\over 16}\, \alpha^4
& + {1005\over 128}\, \alpha^5
& + {1927\over 256}\, \alpha^6
\\
\\
\omega^{(iv)}(k_0)
& =
& 
& + 3\, \alpha^2
& + 3\, \alpha^3
& + {21\over 8}\, \alpha^4
& + {177\over 16}\, \alpha^5
& + {1557\over 128}\, \alpha^6
\end{array}
\ .
\label{wderivs_lad}
\end{equation}

\newpage
\begin{figure}
$$\epsfxsize=4truein\epsffile{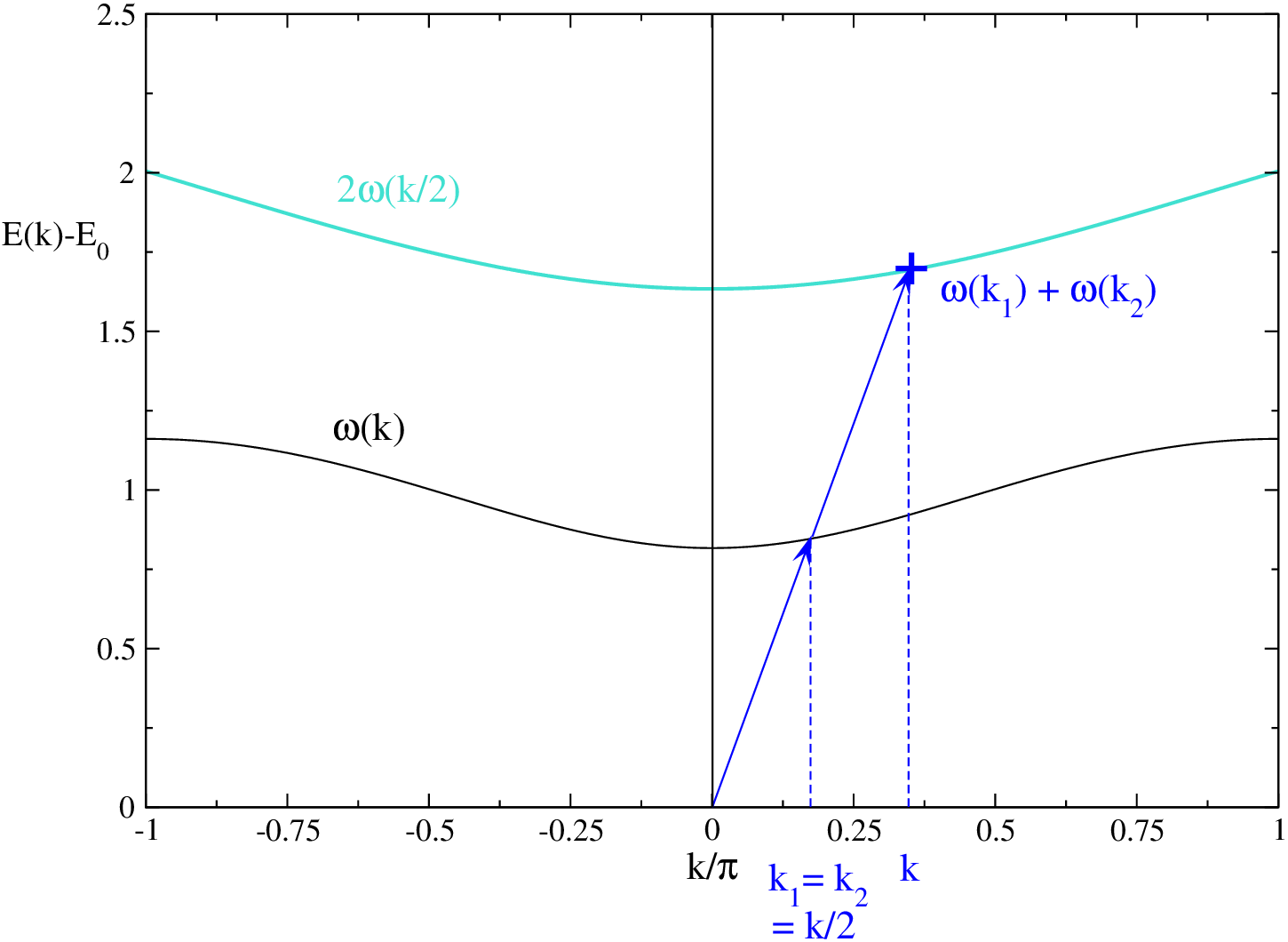}$$
\center{Figure~1.
The alternating chain one-magnon dispersion relation 
$\omega(k)$ for $J=1$ and $\alpha=0.3$, showing a 
symmetric two-magnon state.} 
\end{figure}

\newpage
\begin{figure}
$$\epsfxsize=4truein\epsffile{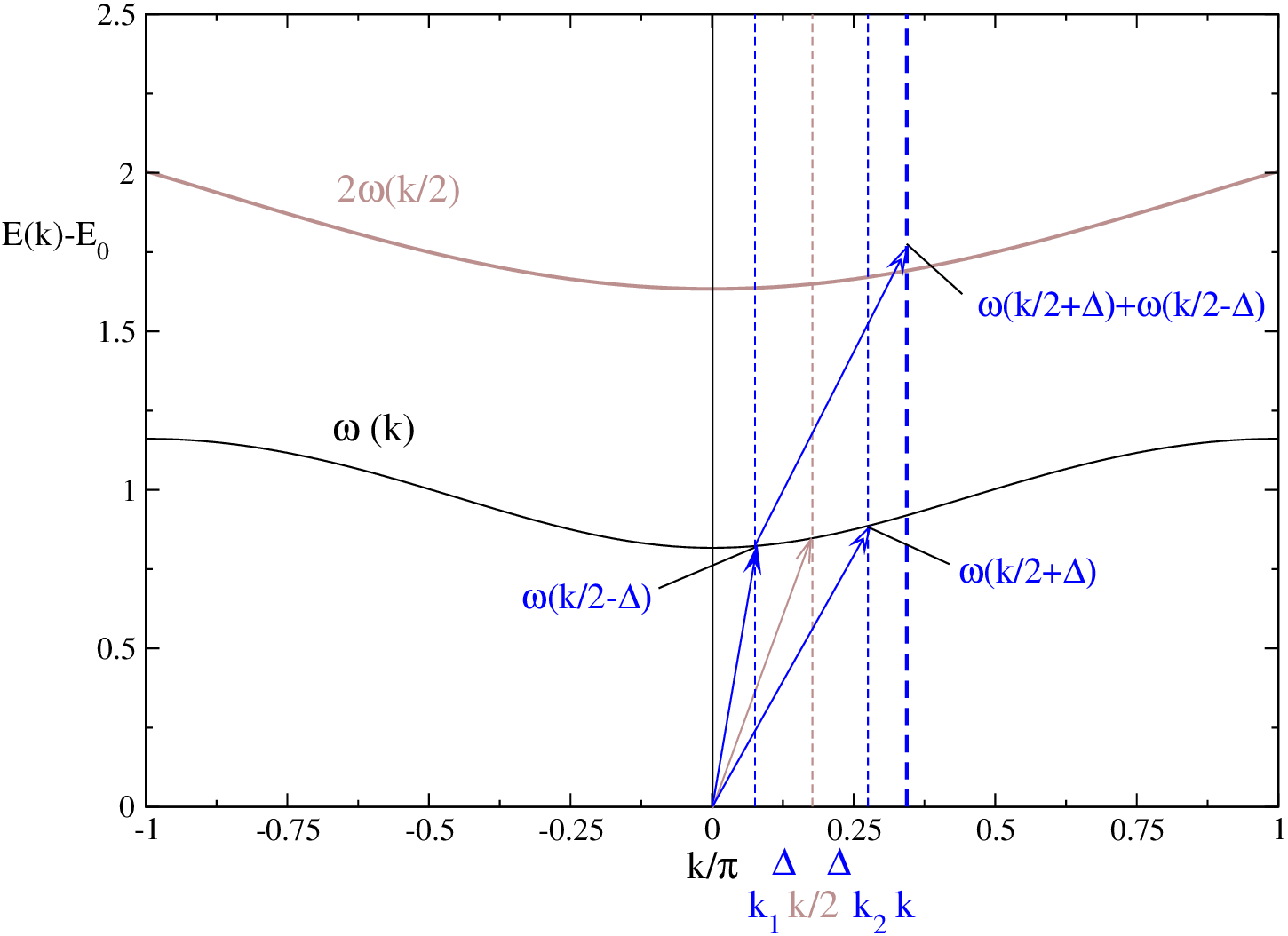}$$
\center{Figure~2.
Illustrating the construction of an asymmetric 
two-magnon state (4).} 
\end{figure}

\newpage
\begin{figure}
$$\epsfxsize=4truein\epsffile{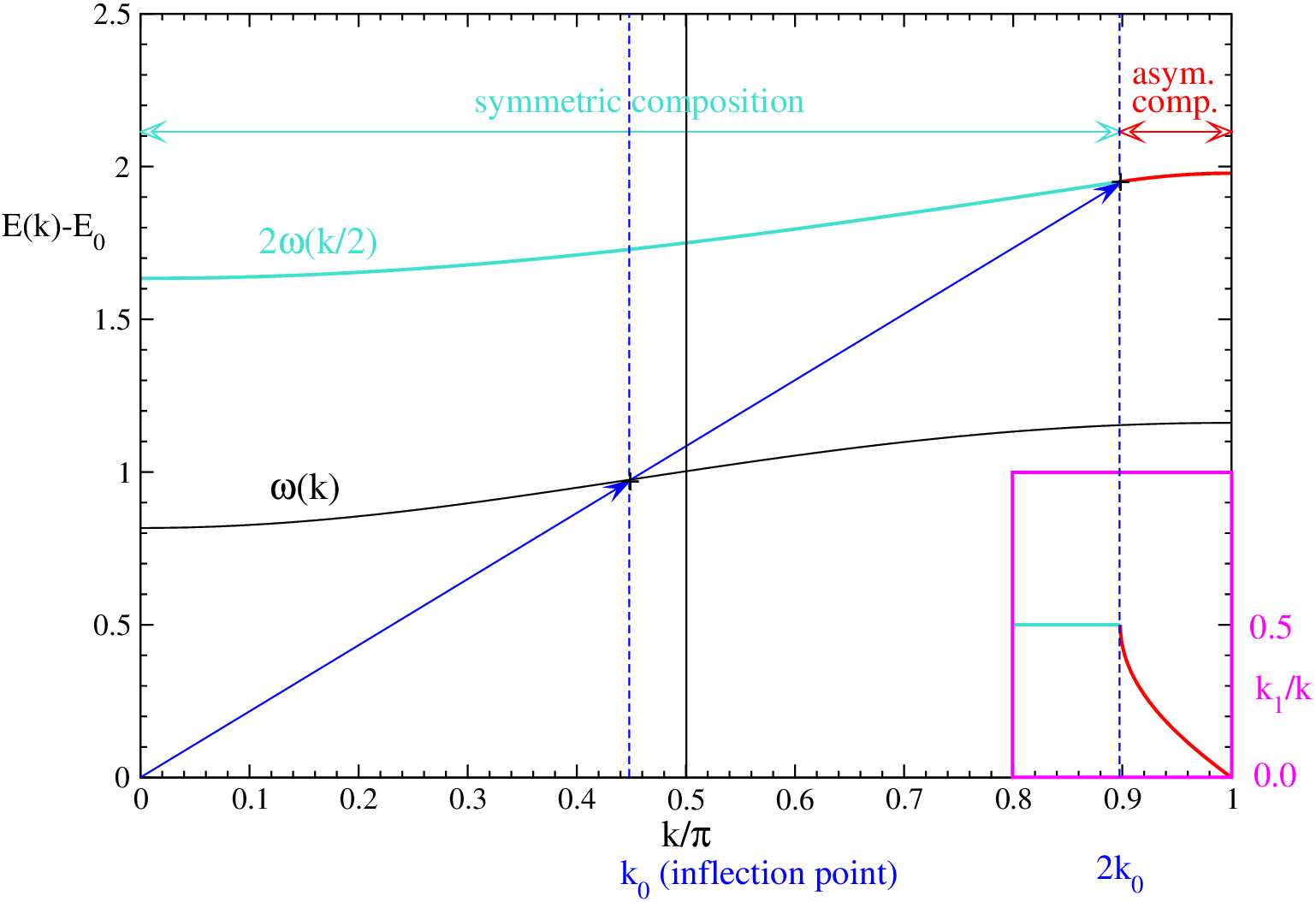}$$
\center{Figure~3.
The $k=2k_0$ transition from symmetric ($k_1=k_2$) to asymmetric ($k_1<k_2$) 
states
on the lower boundary $\Omega_2^-(k)$ of the two-magnon continuum. 
(Alternating chain, $J=1$ and $\alpha=0.3$; $2k_0/\pi =0.898$,
$\Omega_2^-(2k_0)=1.950$.)
The composition break in $k_1/(k_1+k_2)$ is shown at lower right.}
\end{figure}

\begin{figure}
$$\epsfxsize=4truein\epsffile{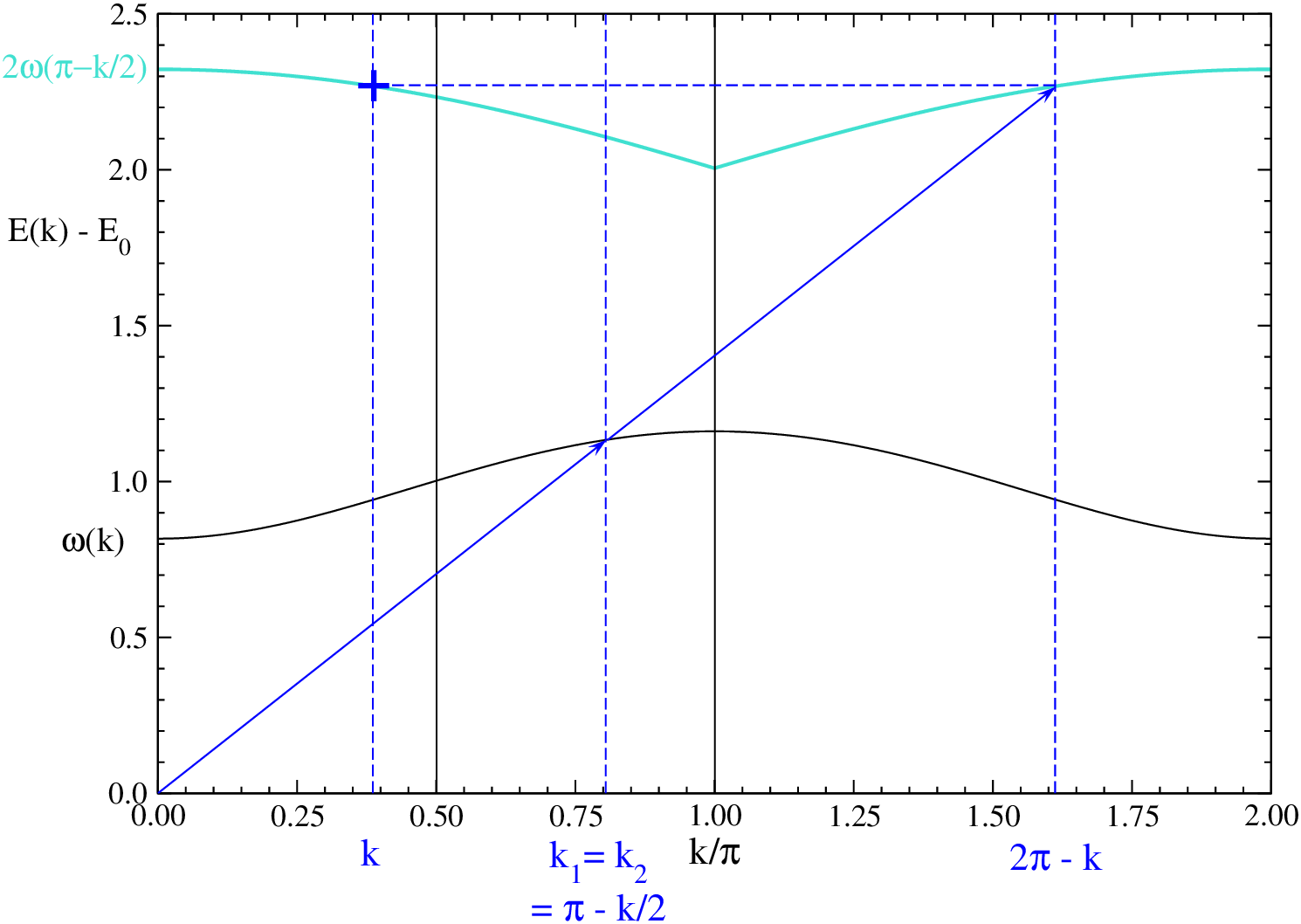}$$
\center{Figure~4.
Construction of the two-magnon upper boundary $\Omega_2^+(k)$. 
(Alternating chain, $J=1$ and $\alpha=0.3$.)}
\end{figure}

\begin{figure}
$$\epsfxsize=4truein\epsffile{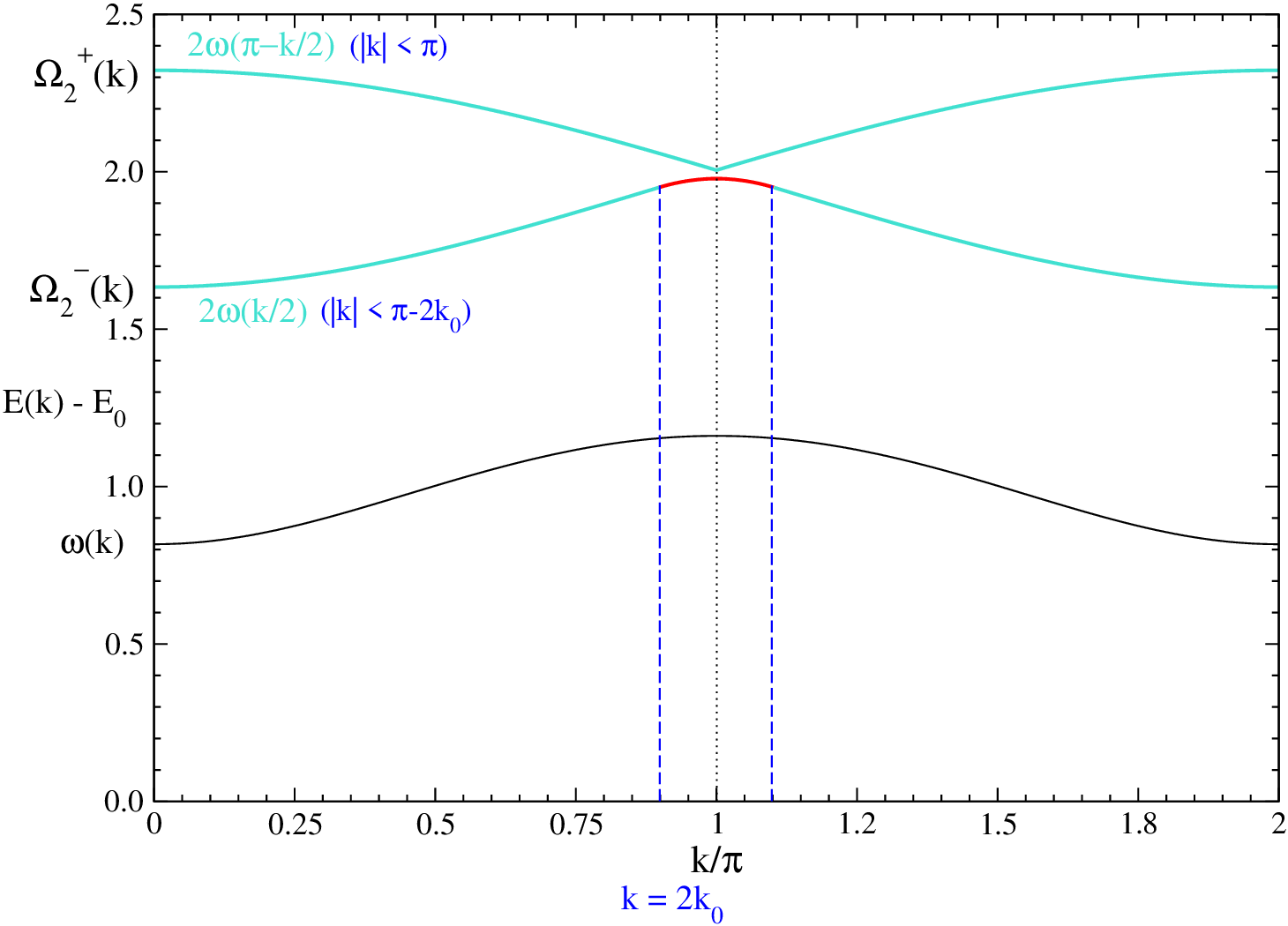}$$
\center{Figure~5.
The complete two-magnon continuum 
boundary $\Omega_2^+(k)$ and $\Omega_2^-(k)$,
showing the composition break at $k=2k_0$. 
(Alternating chain, $J=1$ and $\alpha=0.3$.)}
\end{figure}

\begin{figure}
$$\epsfxsize=4truein\epsffile{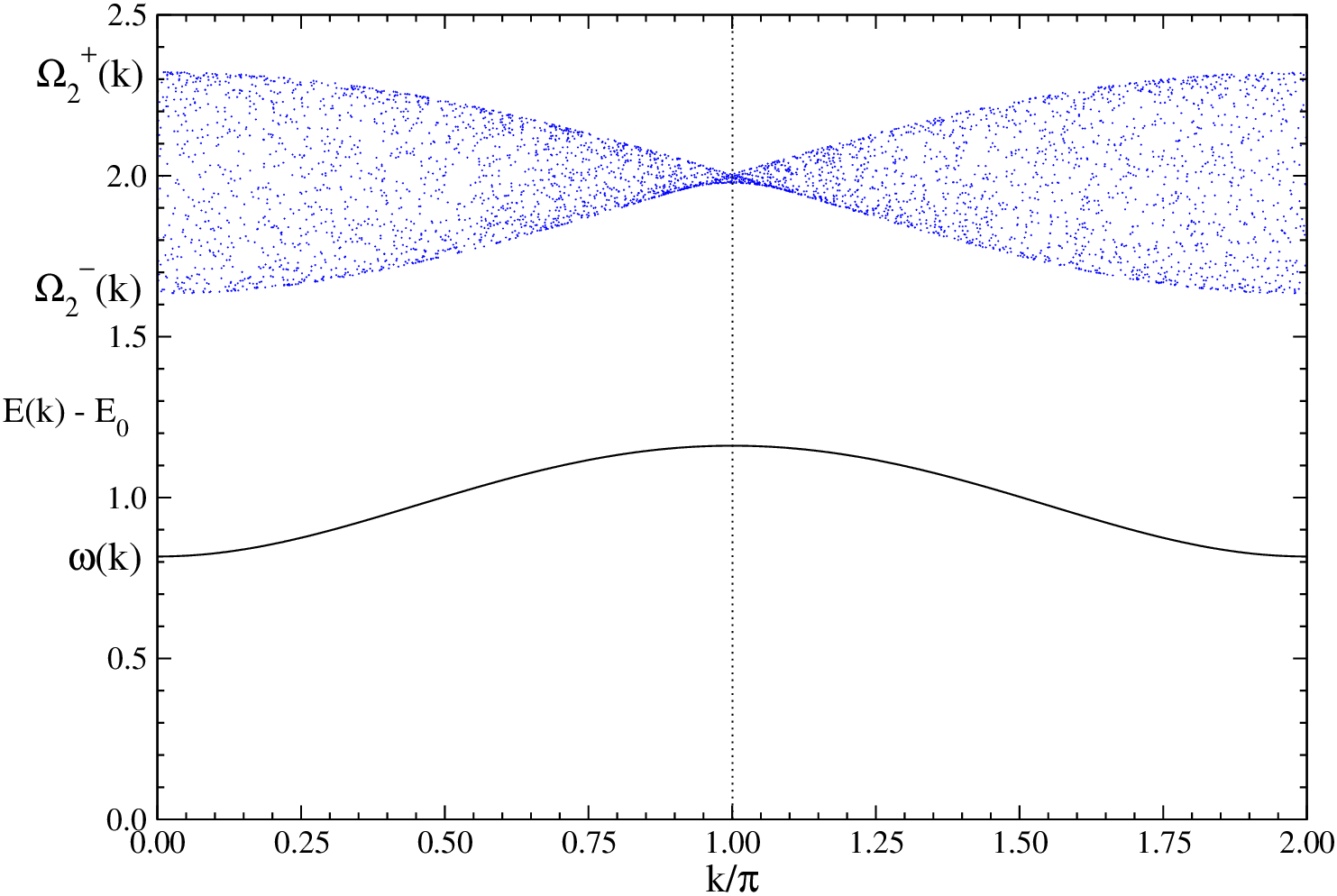}$$
\center{Figure~6.
The density of two-magnon states within
the continuum.
(Alternating chain, $J=1$ and $\alpha=0.3$; flat distribution
in $k_1$ and $k_2$ of $2^{12}$ states.)}
\end{figure}

\begin{figure}
$$\epsfxsize=4truein\epsffile{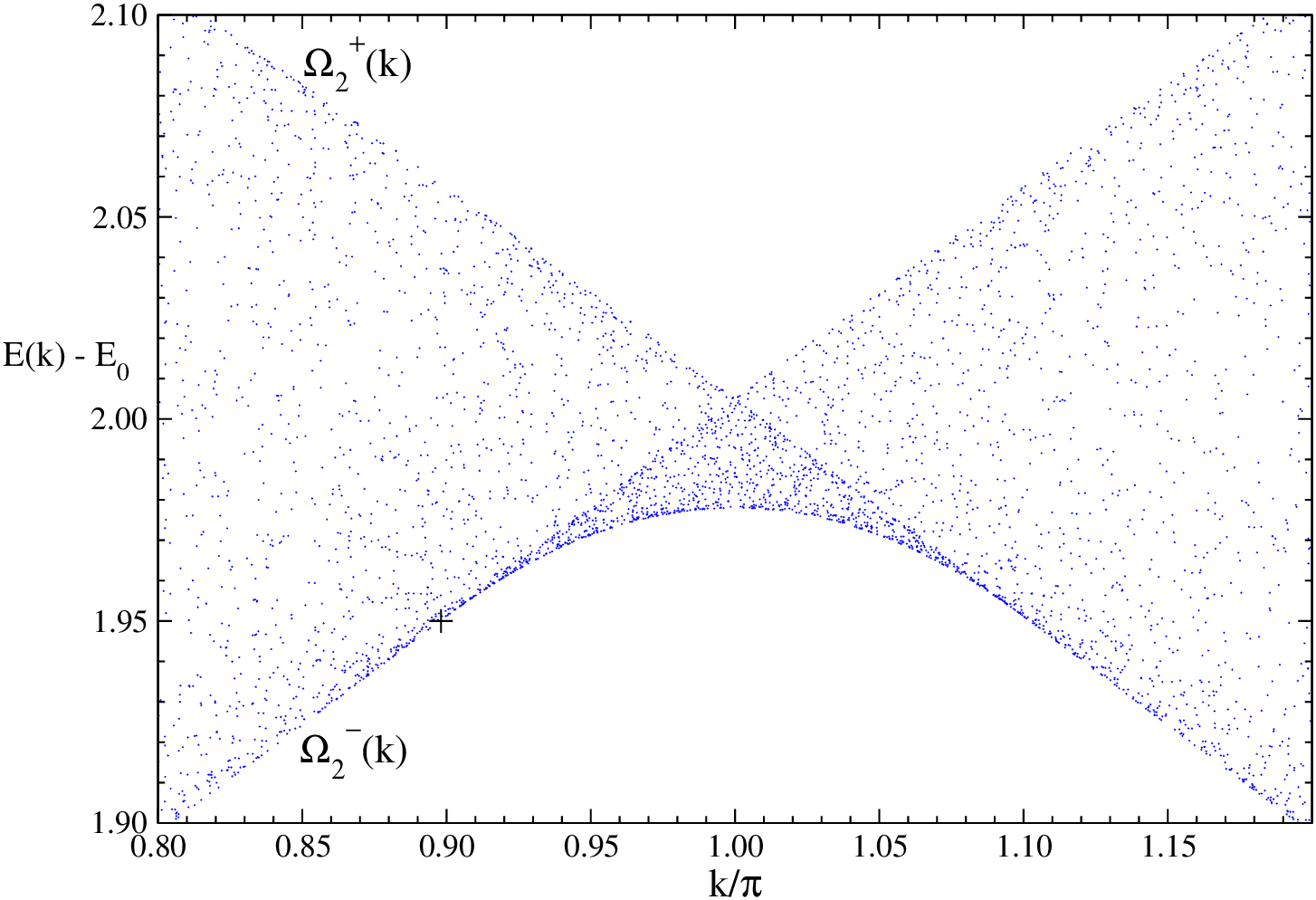}$$
\center{Figure~7.
An enlargement of the alternating chain continuum 
of Fig.6, showing a discontinuity in the density of states
$\rho(k,E)$ (a ``caustic") within the two-magnon continuum. The caustic 
is the line
$2\omega(k/2)$ and its reflection about $k=\pi$. The composition break at $k=2k_0$ where
this curve departs from the lower continuum boundary $\Omega_2^-(k)$ is also
indicated. (Flat distribution in $k_1$ and $k_2$ of $2^{12}$ states
in the region displayed.)}
\end{figure}

\begin{figure}
$$\epsfxsize=4truein\epsffile{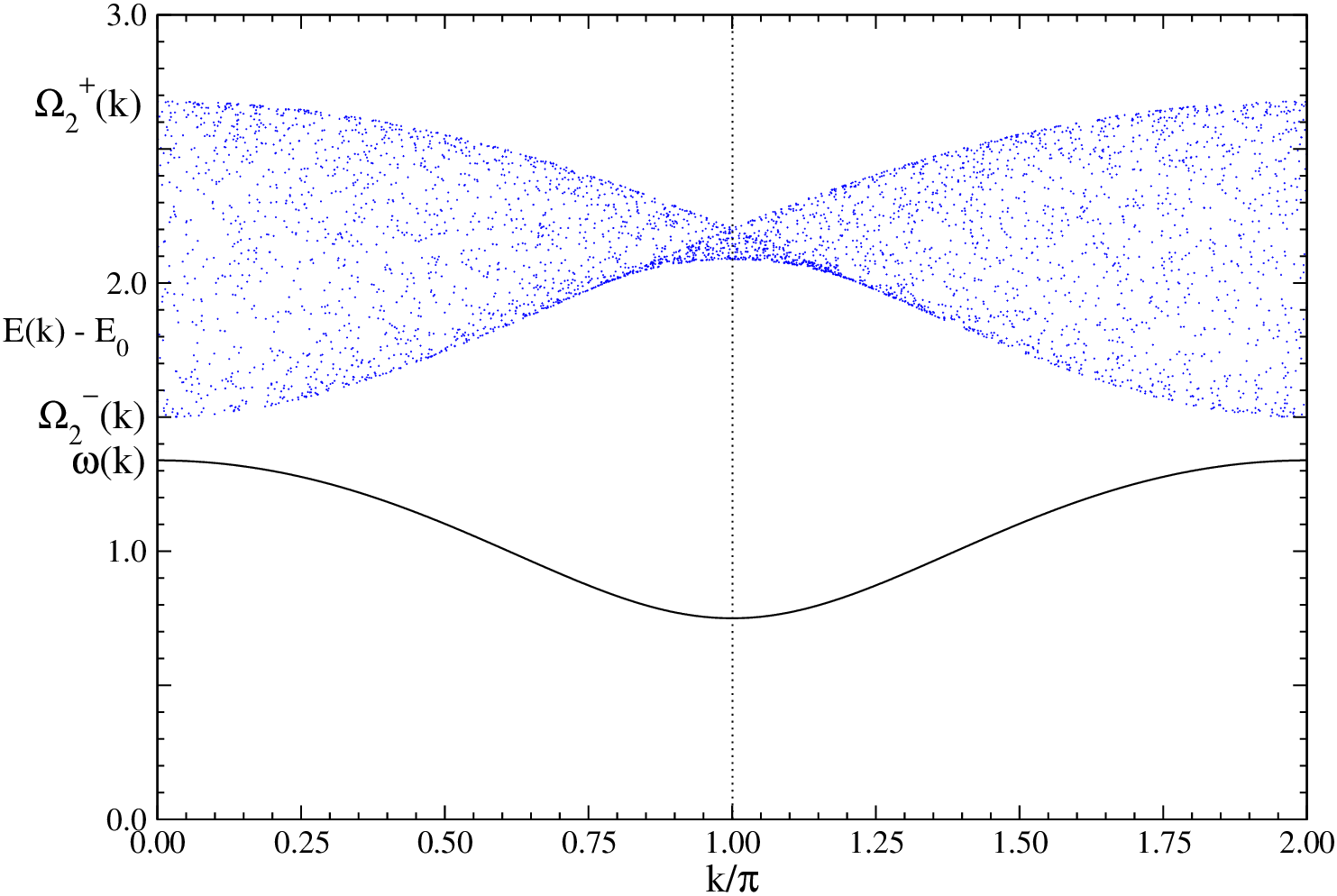}$$
\center{Figure~8.
The one-magnon $\omega(k)$ and density of states within the 
two-magnon 
continuum for a two-leg spin ladder.
($J_{\perp}=1$ and $\alpha=0.3$; flat distribution
in $k_1$ and $k_2$ of $2^{12}$ states.)}
\end{figure}
\vfill\eject

\begin{figure}
$$\epsfxsize=4truein\epsffile{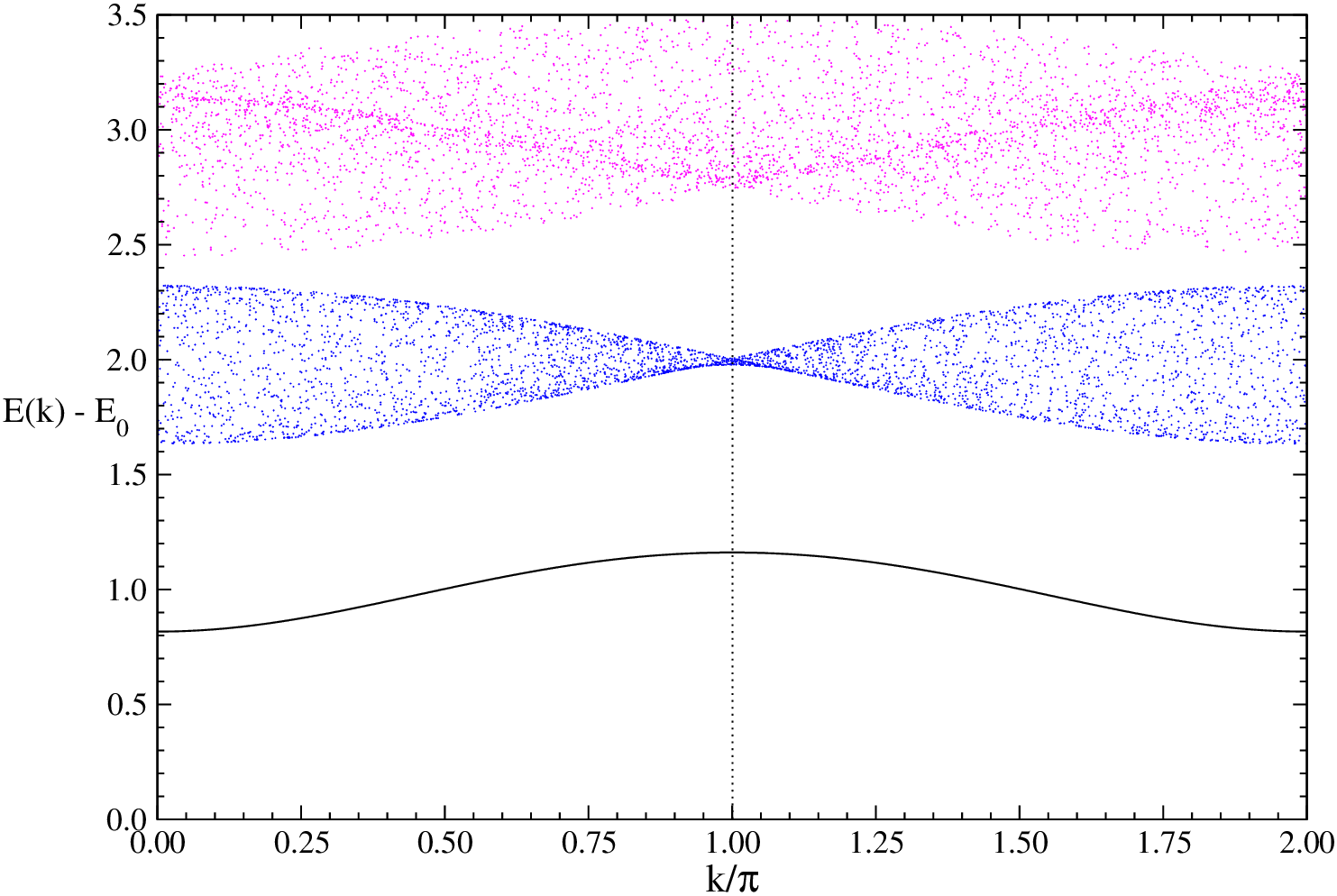}$$
\center{Figure~9.
Two- and three-magnon continua of the alternating chain.
($J=1$ and $\alpha=0.3$; flat distribution
in $k_1$ and $k_2$ of $2^{12}$ states in each case.)}
\end{figure}

\begin{figure}
$$\epsfxsize=4truein\epsffile{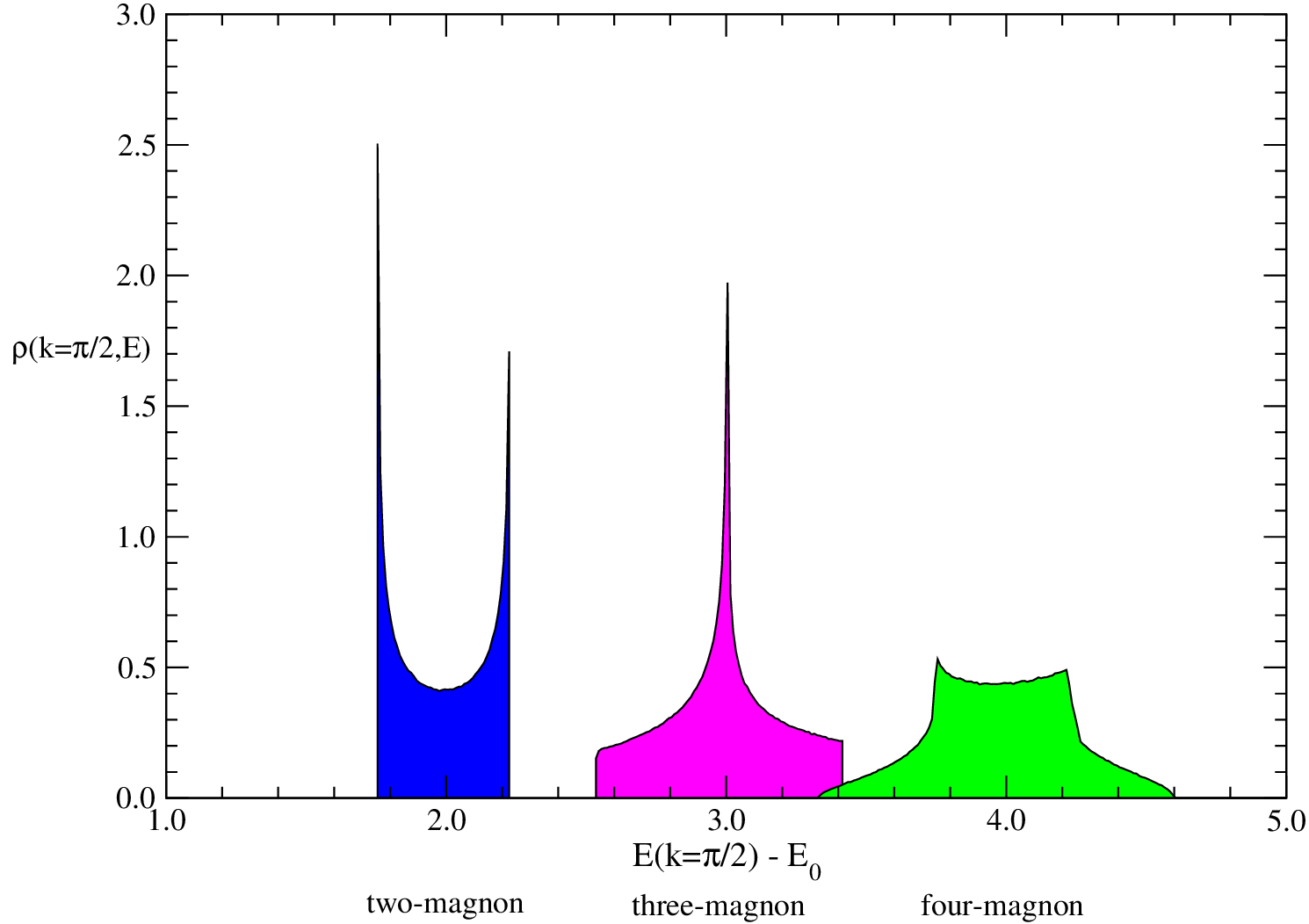}$$
\center{Figure~10.
The density of states $\rho(k,E)$ encountered in a fixed-$k$
slice $(k=\pi/2)$ through the two-, three- and four-magnon continua.
(Alternating chain, $J=1$ and $\alpha=0.3$.)}
\end{figure}

\end{document}